\documentstyle[12pt,epsf]{article}
\newcommand{\be}{\begin{equation}}
\newcommand{\ee}{\end{equation}}
\begin{document}
\title{Quantum Vibrational Impurity Embedded in a One-dimensional Chain
\footnote{submitted to Phys. Lett. A}}
\author{M. I. Molina$^{\dagger}$, J. A. R\"{o}ssler$^{\dagger}$ and G. P. Tsironis$^{*}$\smallskip\  \\
\\
{\it \ }$^{\dagger}${\it \ Departamento de F\'{\i}sica, Facultad de Ciencias}\\
{\it \  Universidad de Chile}\\
{\it \ Casilla 653, Santiago, Chile}\\
%\smallskip\ \\
{\it \ }$^{*}${\it Department of Physics, University of Crete}\\
{\it \ and Research Center of Crete}\\
{\it \ P.O. Box 2208,\ 71003 Heraklion-Crete, Greece}\\
%{\it \ }\smallskip\ \\
{\it \ }}
\date{}
\maketitle

\begin{center}
{\bf Abstract}
\end{center}

We perform a fully quantum mechanical numerical calculation for the problem 
of a single electron (or excitation) propagating in a $N$-site 
one-dimensional chain in the
presence of a single Holstein impurity.  We compute the long-time averaged  
probability for finding the  electron on the impurity site as a function
of the nonlinearity parameter, defined in terms of the electron-phonon
coupling strength and the oscillator frequency.
The results, in the intermediate nonlinearity 
parameter range, differ  substantially from the ones obtained through the 
use of the discrete nonlinear Schr\"{o}dinger equation, even in the
high-frequency regime.

\newpage 

Interest concerning issues related to the interplay between nonlinearity and 
disorder in  discrete systems lead us to consider some time ago, 
the problem of a
single nonlinear (cubic power) impurity embedded in a discrete, linear, 
one-dimensional 
lattice. By using the lattice Green functions in a selfconsistent manner  
we were able to derive analytically all the  bound state properties and the
transmission coefficient through the impurity\cite{mt_prb,mt_ijmp}. We found
an impurity nonlinearity threshold below which, there is no bound state, in marked
contrast with the linear impurity case\cite{economou}. The 
dynamics of the problem showed the presence of a selftrapping transition, when 
the impurity nonlinearity exceeded a threshold
value\cite{mt_ijmp,cmt_jpcm,dkr_pra}.
We also considered the case of a generalized impurity and, by using either an 
appropriate {\em ansatz}\cite{hmt_pre} or the Green function 
formalism\cite{m_ift}, were able
to construct analytically a phase diagram in parameter space showing  
regions with different number of bound states. Extensions of the nonlinear 
impurity problem to a dimerized chain and to higher dimensions have been 
recently investigated\cite{new}.

In all of the above studies, the evolution of a quasiparticle propagating in
the lattice is given by the discrete nonlinear Schr\"{o}dinger equation
(DNLS):
\be
i {{dC_n}\over{dt}} = 
	V (C_{n-1} + C_{n+1}) - \chi |C_{n}|^{\sigma -1}\ C_{n}\ \delta_{n0}\label{eq:1}
\ee
where $C_{n}(t)$ denotes the probability amplitude for the quasiparticle
to be at site $n$, $V$ is the nearest-neighbor transfer energy integral,
$\chi$ is the nonlinearity parameter and $\sigma$ is the nonlinearity
exponent. In the Holstein impurity context,
the DNLS (for the cubic nonlinearity, $\sigma = 3$) describes the time evolution 
of an electron which is 
strongly coupled to the vibrational degrees of freedom of a molecule at the impurity 
site.
In this case $\chi = g^{2}/\omega$ where $g$ is the electron-phonon coupling and
$\omega$ is the phonon frequency.

However, several doubts have been raised concerning the general
applicability of the 
semiclassical approximation on which the DNLS is 
based to the fully quantum mechanical Holstein
problem[9 - 13]. This has motivated
us to calculate the effect of quantum fluctuations in these impurity systems.
In the present letter we perform a fully quantum mechanical 
numerical calculation
for a single vibrational impurity coupled to an electron (or excitation) at a 
given site ($j=0$) of a finite chain with periodic boundary conditions
and compare the results to the DNLS 
predictions. The Hamiltonian is
\be
H \; = \; V \sum_{j=0}^{N-1}\{ C_{j}^{\dagger} C_{j+1} + C_{j+1}^{\dagger} C_{j} \}
+ \omega a^{\dagger} a + g (a + a^{\dagger})\;C_{0}^{\dagger} C_{0}
	\label{eq:2}
\ee
where $C_{j}^{\dagger}(C_{j})$ creates (destroys) an electron on the $jth$-site,
$a^{\dagger}(a)$ is the usual phonon creation (destruction) operator, 
$\omega$ is the (Einstein) oscillator frequency and $g$ is the electron-phonon 
coupling at the impurity site. The most general eigenfunction has the form
\be
|\Psi_{E}\rangle \; =
\; \sum_{j=0}^{N-1}\sum_{\nu=0}^{\infty} \alpha_{j}^{(\nu)} |\nu\rangle\;\otimes\;| j\rangle
			\label{eq:3}
\ee
where
$|\nu\rangle= (a^{\dagger})^{\nu}/\sqrt{\nu!}\;|0\rangle_{ph}$
and $|j\rangle = C_{j}^{\dagger}\ |0\rangle_{e}$. The states $|0\rangle_{ph}$ and
$|0\rangle_{e}$ are the phonon and electron vacuum, respectively. After inserting
Eq.(\ref{eq:3}) into the eigenvalue equation $H |\Psi_{E}\rangle = E\ |\Psi_{E}\rangle$,
we obtain

\[
\sum_{j=0}^{N-1}\sum_{\nu=0}^{\infty} [\ E - \omega \nu\ ]\ \alpha_{j}^{(\nu)}
C_{j}^{\dagger} |\nu\rangle = V \sum_{j=0}^{N-1} \sum_{\nu=0}^{\infty}\ [\ 
\alpha_{j-1}^{(\nu)} + \alpha_{j+1}^{(\nu)} \ ] C_{j}^{\dagger} |\nu\rangle 
\]
\be
\ \ \ \ \ \ \ \ \ \ \ \ \ \ \ \ \ \ \ \ + g\ C_{0}^{\dagger} \sum_{\nu=0}^{\infty}
[ \ \alpha_{0}^{(\nu +1)} \sqrt{\nu +1} + \alpha_{0}^{(\nu -1)} \sqrt{\nu} \ ] |\nu\rangle
\label{eq:4}
\ee
The use of the transfer matrix formalism and periodic boundary conditions
allows us 
to express all the $\alpha_{j}^{(\nu)}$ in terms of the
$\alpha_{0}^{(\nu)}$ viz.,
\be
\alpha_{j}^{(\nu)} = {\alpha_{0}^{(\nu)} \over{\sin(N k_{\nu})}}[ \ \sin(j k_{\nu}) +
\sin((N-j) k_{\nu}) \ ] \label{eq:5}
\ee
where the wavevector $k_{\nu}$ is defined by
$A_{\nu}\equiv (E - \omega\ \nu)/2 V = \cos(k_{\nu})$
and the eigenvalue equation reduces to
\be
(g/V) \sin(N k_{\nu})\ \{ \alpha_{0}^{(\nu +1)}\ \sqrt{\nu +1} \ +\  \alpha_{0}^{(\nu -1)}
\sqrt{\nu} \}\ =\ 2\ \alpha_{0}^{(\nu)}\ \sin(k_{\nu})\ [\ \cos(N k_{\nu}) -1\ ]
\label{eq:6}
\ee
%%%%%%%%%%%%%%%%%%%%%%%%%%%%%%%%%%%%%%%%%%%%%%%%%%%%%%%%%%%%%%

When $|A_{\nu}| > 1$, $k_{\nu}\rightarrow i\ K_{\nu} + \{0,\pi\}$
%\{ \begin{array}{c}
%0\\\pi
%\end{array} \}$
where the left (right) choice corresponds to
$A_{\nu} > 1\ (A_{\nu} < -1) $ and where $K_{\nu}$ is given by
$K_{\nu} = \log( |A_{\nu}| + \sqrt{A_{\nu}^{2} - 1} )$. The
trigonometrical equations (\ref{eq:5}), (\ref{eq:6}) change to
a hyperbolic form:
\be
\alpha_{j}^{(\nu)} = {\alpha_{0}^{(\nu)} \over{\sinh(N K_{\nu})}}[\ 
(\pm 1)^{N}\ \sinh(j K_{\nu}) + \sinh((N-j) K_{\nu}) \ ] \ (\pm 1)^{j}
\ee
\be
(g/V) \sinh(N K_{\nu})\{ \alpha_{0}^{(\nu +1)}\ \sqrt{\nu +1} \ +\  \alpha_{0}^{(\nu -1)}
\sqrt{\nu} \} = \pm 2\ \alpha_{0}^{(\nu)} \sinh(K_{\nu})\{\cosh(N K_{\nu}) - (\pm 1)^{N} \}
\label{eq:7}
\ee
where the $\pm$ sign corresponds to the sign of $A_{\nu}$.
%%%%%%%%%%%%%%%%%%%%%%%%%%%%%%%%%%%%%%%%%%%%%%%%%%%%%%%%%%%%%%%%%%%%%%%%%
Equations (\ref{eq:6}) and (\ref{eq:7}) constitute an infinite set of
``three-terms'' recursion relations of the form:
\be
c_{\nu}\ \alpha^{(\nu -1)} + b_{\nu} \ \alpha^{(\nu)} + a_{\nu} \ \alpha^{(\nu+1)} = 0
\label{eq:9}
\ee
The eigenvalue equation associated to this (infinite) tridiagonal equation
can be solved by a standard method. However, some care must be taken
in order to avoid numerical instabilities:\ we iterate Eq.(\ref{eq:9})
from two directions, ``up'' from $\nu=0$ up to $L$  and ``down'' from 
$\nu_{max}$ down to $L$, where $\nu_{max}$ is an appropriate phonon 
number cutoff. The eigenvalue equation is then obtained by imposing that
both iterations yield 
the same value for $\rule[0.1cm]{0cm}{0.4cm}\alpha_{0}^{(L + 1)}/\alpha_{0}^{(L)}$ at the 
``gluing'' point $\nu = L$, chosen to ensure the numerical stability of the
iteration process. The precision of the numerical computation is monitored 
through several tests:\\
(a)\ \ Evaluation of 
\be
\left|\  \left\{{\alpha_{0}^{(L+1)}\over{\alpha_{0}^{(L)}}}\right\}_{up} -
\left\{{\alpha_{0}^{(L + 1)}\over{\alpha_{0}^{(L)}}}\right\}_{down} \right| +
\left|\  \left\{{\alpha_{0}^{(L+2)}\over{\alpha_{0}^{(L+1)}}}\right\}_{up} -
\left\{{\alpha_{0}^{(L + 2)}\over{\alpha_{0}^{(L+1)}}}\right\}_{down} \right|
\ee
(b)\ \ Orthogonality relations:\ for instance,
$\sum_{\mu} \alpha_{0,\mu}^{(0)}\ \alpha_{0,\mu}^{(\nu)} = 
\delta_{\nu,0}$ where the index $\mu$ runs over all the eigenstates.\\
(c)\ \ A general test based on the Hamiltonian: If $A$ is an arbitrary operator, then
the mean value of $\Omega\equiv [A, H]$ over any eigenstate of $H$ is 
identically zero. For instance, if we take $A = a$, the phonon destruction 
operator,
the identity $ 0 = \langle \Omega \rangle $ leads to the condition
$\omega < a > \ +\  g < C_{0}^{\dagger} C_{0} > = 0$.

Once in possession of all the eigenenergies $E_{\mu}$ and the 
amplitudes $\alpha_{j,\mu}^{(\nu)}$, we are ready to
compute dynamical observables. As the initial condition for the electron
we use the one that places it completely on the impurity site at $t=0$
and focus on the probability 
$P_{0}(t)$ for finding it there at an arbitrary time $t$ later. For the
phonon part we use two different initial conditions: an undisplaced oscillator
(zero phonons present) and a naturally ``relaxed'' oscillator (a coherent
state). Due to absence of degeneration, the long-time average probability
$< P_{0} > = \lim_{T\rightarrow \infty} \ (1/T)\ \int_{0}^{T} P_{0}(t)\ dt$
can then be expressed
in terms of the $\alpha_{0,\mu}^{(\nu)}$ as
\[
\mbox{Initial state:}\hspace{1cm}|\Psi(0)> =  | j=0 > \otimes
\sum_{n=0}^{\infty} p_{n} | n > \nonumber
\]
\be
\mbox{and as a result}\hspace{1cm}< P_{0} >\ =\ \sum_{\mu}\ [\sum_{\nu=0}^{\infty}\ 
{\alpha_{0,\ \mu}^{(\nu)}}^{2}\ ]
 \ [ \ \sum_{n=0}^{\infty}\ p_{n}\ \alpha_{0,\ \mu}^{(n)}\ ]^{2}\label{eq:P0}
\ee
where
\be
p_{n} = \left\{ \begin{array}{cc}
	\delta_{n,0} & \mbox{undisplaced oscillator} \\
	{\rule[-0.7cm]{0cm}{0.7cm}{\displaystyle {1\over{\sqrt{n!}}} \left({-g\over{\omega}}\right)^{n}} 
		\exp[-(g/\omega)^{2}/2]}
		     & \mbox{coherent state case}
	\end{array}
	\right.
\ee

We have examined the behavior of Eq.(\ref{eq:P0}) as a function of the
nonlinearity parameter $\chi/V \equiv g^{2}/\omega V$, for different
oscillator frequencies ($\omega \propto 1/\sqrt{\mbox{mass}}$), for chains with $N = 2,3,4$ and $5$, the
chain corresponding to $N=2$ being the only nonperiodic one.
We show the results for $N=2$ and $N= 4$ in figures 1-4 where we also
compare with the quantity
$< P_{0} >$ evaluated though the DNLS equation. 
We observe significant departure from the DNLS prediction for 
intermediate nonlinearity, even for very large $\omega$. In the case
of an (initially) undisplaced oscillator, while $< P_{0} >$ increases 
with $\chi$ on average, it also shows oscillatory behavior with $\chi$ 
for large oscillator mass. For smaller masses, the oscillatory
behavior decreases setting into a monotonic curve for the smallest mass
examined. This case, which is closer to the antiadiabatic limit where the
DNLS equation should work, 
still differs significantly from the DNLS prediction. 
For the case of an (initially) displaced oscillator, $< P_{0} >$ increases 
always monotonically with the nonlinearity
parameter. In both cases, as the oscillator mass decreases, the curves
converge to a {\em linear}-like  impurity curve\cite{mt_prb,economou},
remaining significantly 
away from the DNLS prediction, down to the lowest mass considered here.
For $\omega = 100 V$ the curve is almost undistinguishable from the one
corresponding to a static, linear impurity of strength $\chi$.
We note, in particular, the dramatic difference in the slopes
of the quantum mechanical versus the DNLS curve in the small
coupling limit. The quantum mechanical impurity results approach
the DNLS impurity results only in the very large coupling limit
$\chi\rightarrow \infty$. Similar
results are found for the $N = 3 $ and $N = 5$ case.

The equivalence of model (\ref{eq:2}) with a static impurity can be
understood as follows: Let us perform the
canonical transformation $H\rightarrow \tilde{H} = U\ H\ U^{-1}$ with
$U = \exp[\ \lambda (a - a^{\dagger}) C_{0}^{\dagger} C_{0} ]$.
With the choice $\lambda = -g/\omega$, the transformed Hamiltonian
becomes:
\be
\tilde{H} = \sum_{j} \ [ \ V_{j} C_{j}^{\dagger} C_{j+1} + h.c. \ ] +
\omega\ a^{\dagger} a - (g^{2}/\omega) \ C_{0}^{\dagger} C_{0}
\ee
where
\be
V_{j} = \left\{ \begin{array}{cc}
	V\ \exp[\ (g/\omega) (a - a^{\dagger})\ ]  & \mbox{$ j = 0 $} \\
	V\ \exp[\ -(g/\omega) (a - a^{\dagger})\ ]  & \mbox{$ j = -1$} \\
	V	& \mbox{$j \neq 0,-1$}
	\end{array}
	\right.
\ee
In the limit $\omega\rightarrow \infty$ with fixed $\chi = g^{2}/\omega$
\be
\tilde{H} \rightarrow \sum_{j} \ [ \ V\ C_{j}^{\dagger} C_{j+1} + h.c. \ ] +
\omega\ a^{\dagger} a - (g^{2}/\omega)\ C_{0}^{\dagger} C_{0}
\ee
which is the Hamiltonian for a chain with a single static, linear
impurity. Therefore, in the limit $\omega\rightarrow \infty$
($\omega\sim 100 V$ is good enough) with fixed
$\chi = g^{2}/\omega$, our Hamiltonian becomes equivalent to the
one of a single, static, linear impurity. The DNLS limit is never
achieved.

We have directly shown in the present work that the problem of a single
vibrational impurity embedded in a finite linear chain, when treated in a completely 
quantum-mechanical framework, does not approach the predictions based on
the DNLS.  We conclude that, for the Holstein
impurity problem the DNLS fails to describe correctly the selftrapping
process. Our results agree with recent studies\cite{salkola} on a
coupled quasiparticle-boson system 
which also finds that the DNLS has a limited range of validity in the
context of this problem. It would be interesting to test the DNLS
in a system possessing translational symmetry where the enhancement of
trapping due to the single impurity character of the present 
model would be absent. Work in that direction is currently under way
by our group.

Finally, It is worthwhile pointing out that
the general solutions for DNLS-like impurity problems that have
been obtained earlier retain their validity and usefulness in classical
problems where DNLS is valid. Examples of these problems are
typically furnished from nonlinear optical applications.

Two of the authors (M. I. M. and J. A. R.) acknowledge partial support from 
FONDECYT grant 1950655 while the third one (G. P. T.) acknowledges support
from $\Pi E N E \Delta $ grant 95-$E \Delta$-115 of the General Secretariat
of Research and Technology of Greece.

\newpage

\centerline{FIGURE CAPTIONS}

\vspace{1.0 cm}

\noindent {\bf Figure 1:}\ $N = 2$ chain: \ Long-time average 
probability of finding the electron (excitation) on the impurity versus
the nonlinearity parameter $\chi/V = g^{2}/\omega V$, for different oscillator
frequencies (masses).\ The oscillator is initially undisplaced.\\

\noindent {\bf Figure 2:}\ \ Same as in figure 1, but for the case of an initially
{\em displaced} oscillator.\\

\noindent {\bf Figure 3:}\ \  $N = 4$ chain: \ Long-time average 
probability of finding the electron (excitation) on the impurity versus
the nonlinearity parameter $\chi/V = g^{2}/\omega V$, for different oscillator
frequencies (masses).\ The oscillator is initially undisplaced.\\

\noindent {\bf Figure 4:}\ \ Same as in figure 3, but for the case of an
initially {\em displaced} oscillator.

\newpage

%%%%%%%%%%%%%%%%%%%%%%  FIGURE 1   %%%%%%%%%%%%%%%%%%%%%%%%%%%%%%%%%%
\begin{figure}[p]
\begin{center}
\leavevmode
\hbox{
\epsfysize=100 pt
\includegraphics{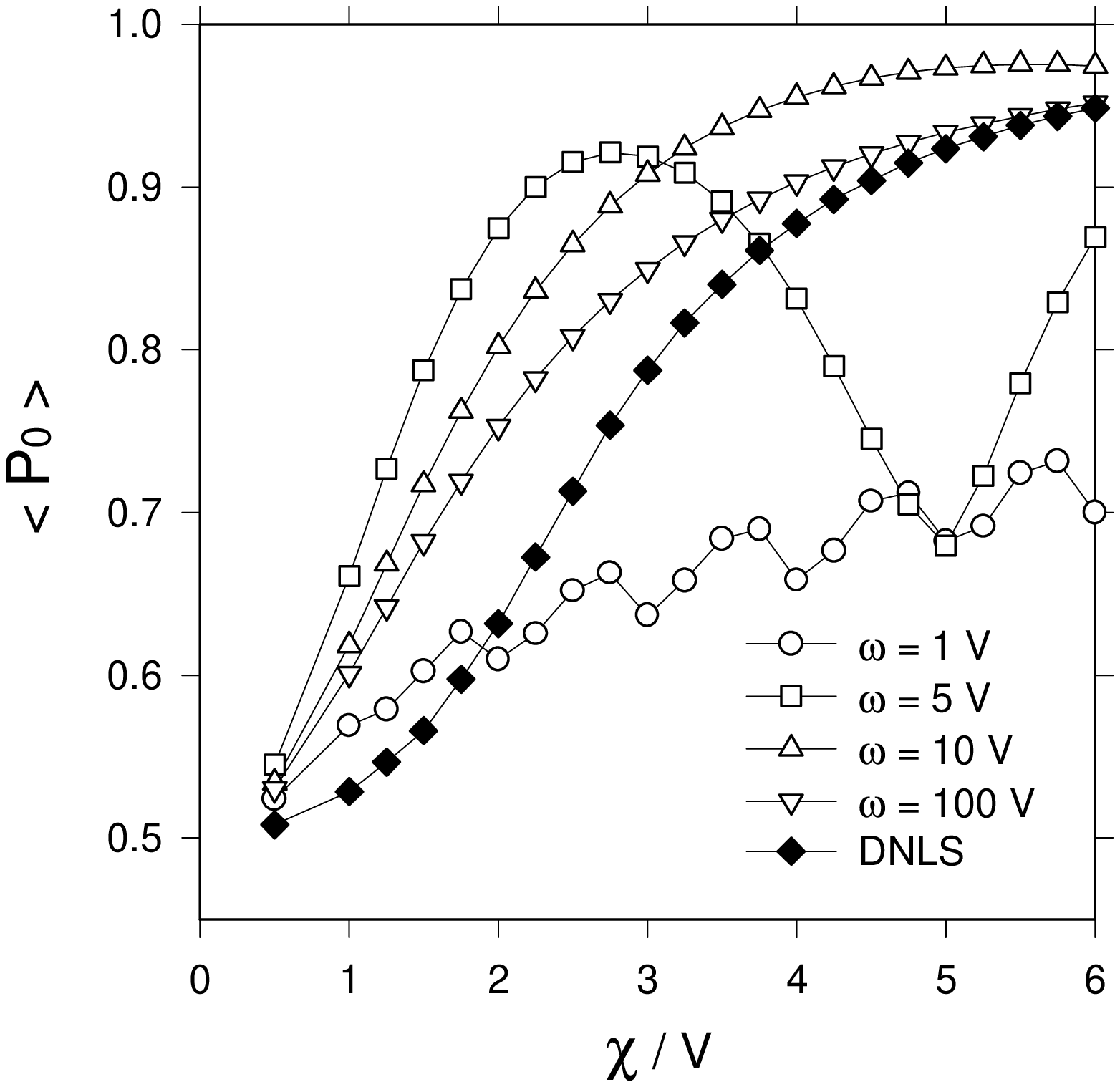}  }
\end{center}
\end{figure}

\clearpage

%%%%%%%%%%%%%%%%%%%%%%%  FIGURE 2   %%%%%%%%%%%%%%%%%%%%%%%%%%%
\begin{figure}[p]
\begin{center}
\leavevmode
\hbox{
\epsfysize=100 pt
\includegraphics{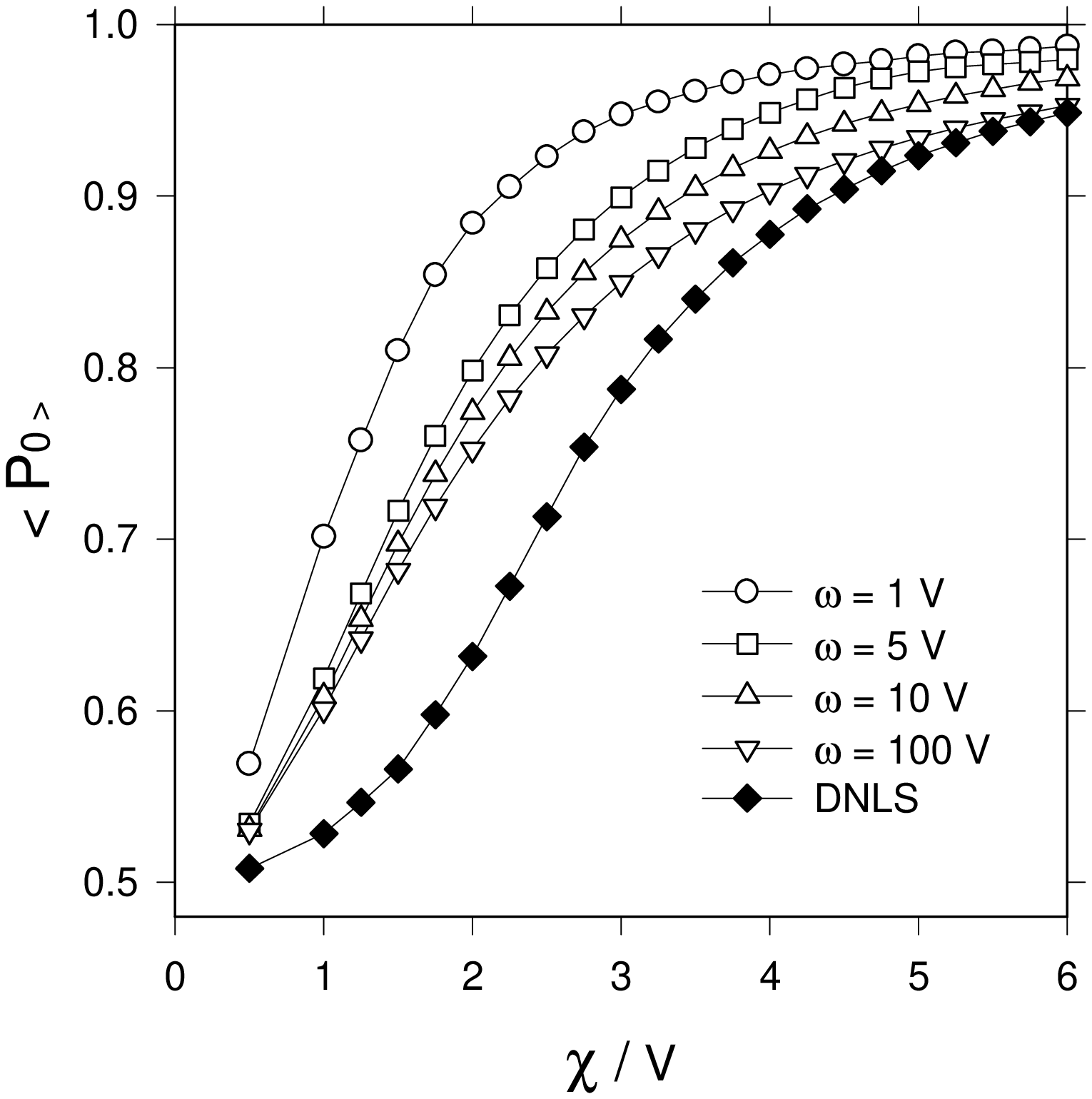}  }
\end{center}
\end{figure}

\clearpage

%%%%%%%%%%%%%%%%%%%%%%%%   FIGURE 3   %%%%%%%%%%%%%%%%%%%%%%%
\begin{figure}[p]
\begin{center}
\leavevmode
\hbox{
\epsfysize=100 pt
\includegraphics{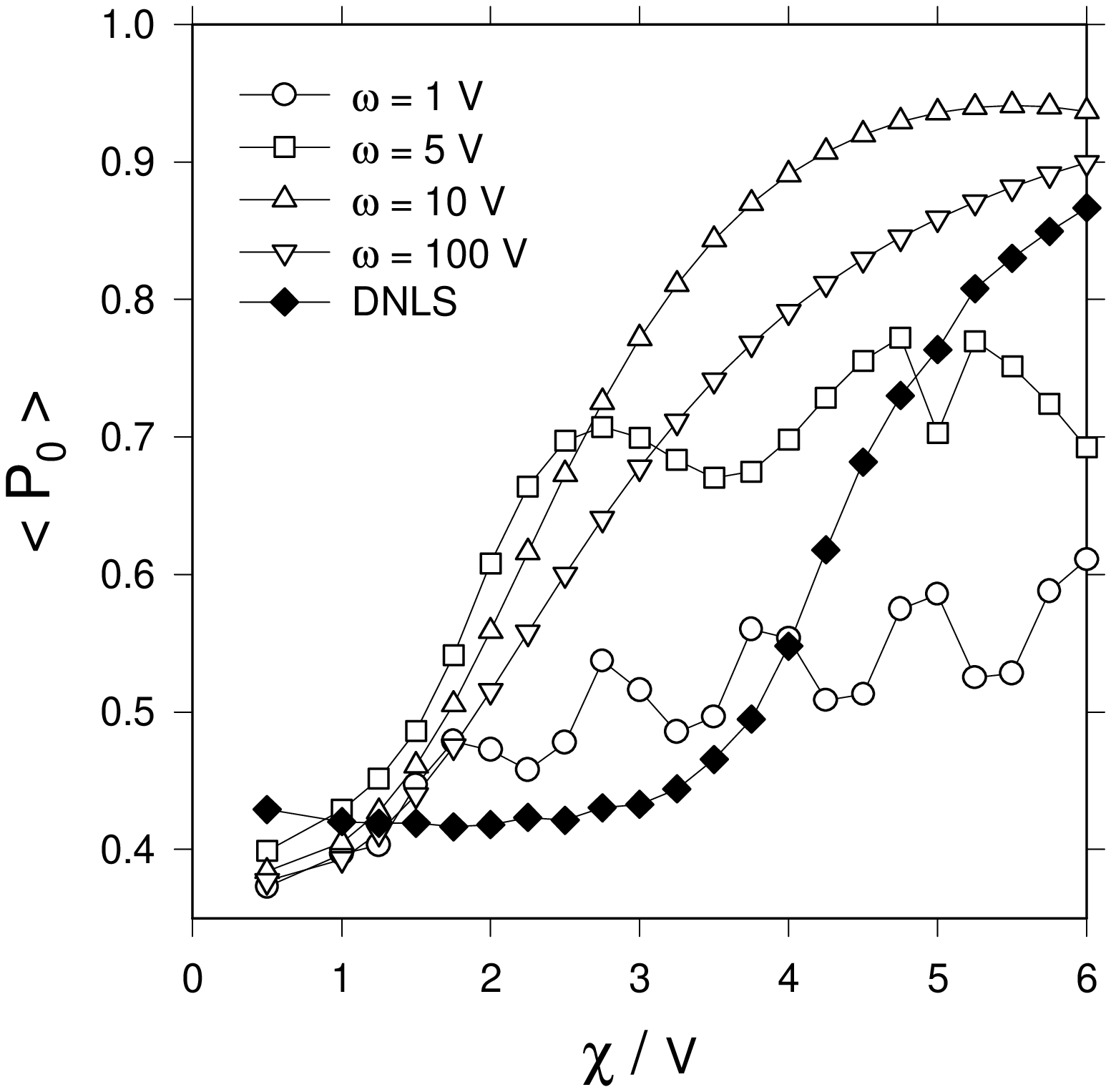}  }
\end{center}
\end{figure}

\clearpage

%%%%%%%%%%%%%%%%%%%%%%%%%   FIGURE 4   %%%%%%%%%%%%%%%%%%%%%%%%%%%
\begin{figure}[p]
\begin{center}
\leavevmode
\hbox{
\epsfysize=100 pt
\includegraphics{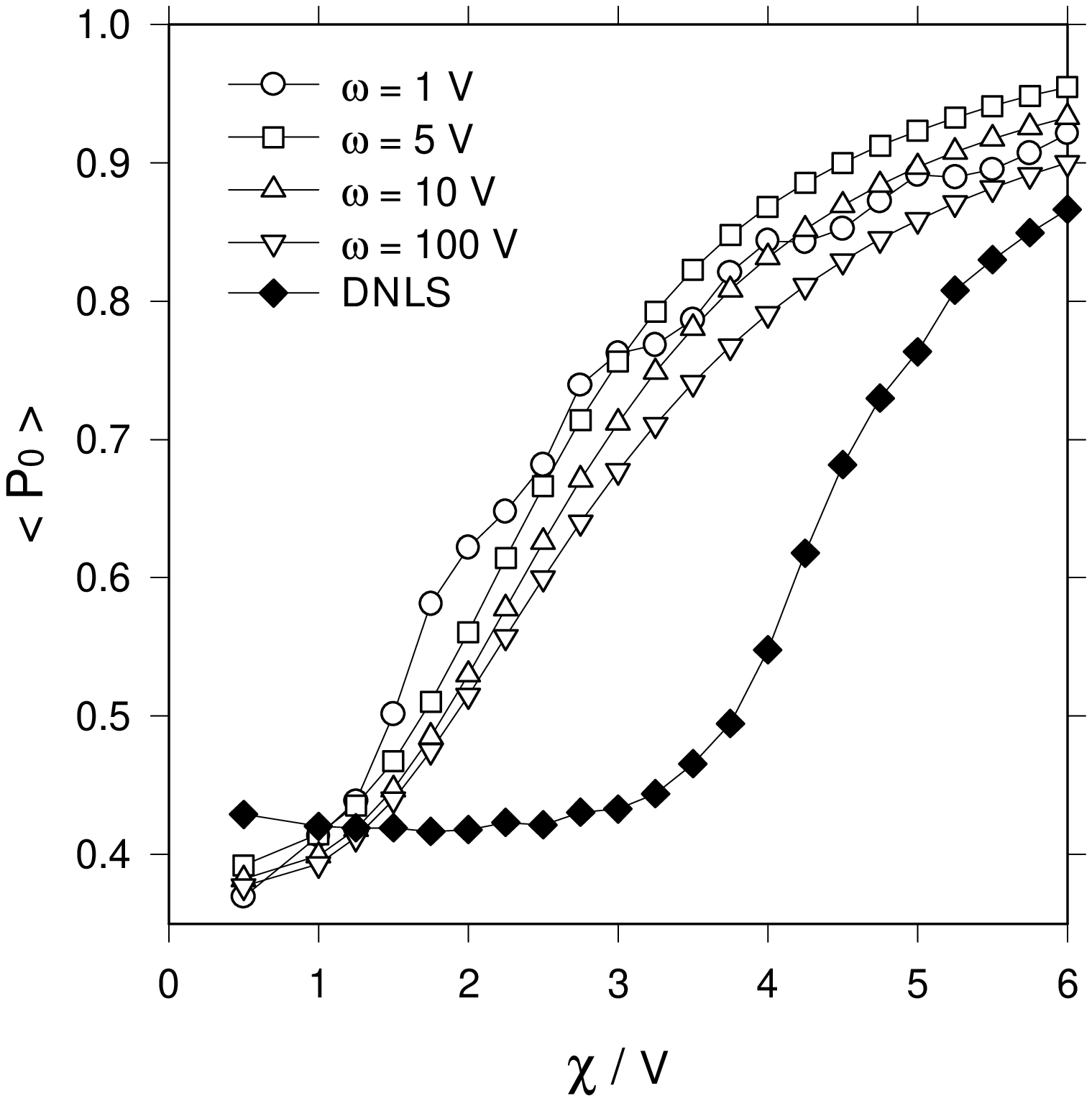}  }
\end{center}
\end{figure}

\end{document}